\documentclass[aps,pra,showpacs]{revtex4}

\usepackage{amssymb}
\usepackage{amsmath}
\usepackage[xdvi,dvips]{graphicx}
\usepackage{graphicx}
\usepackage{subfigure}

\newcommand{\br}{{\bf{r}}}

\begin{document}

\title{Wave patterns generated by a supersonic moving body in a binary
Bose-Einstein condensate}

\author{Yu. G. Gladush$^{1}$, A. M. Kamchatnov$^{1}$, Z. Shi$^{2}$, P. G.
Kevrekidis$^{2}$, D. J. Frantzeskakis$^{3}$, and B. A. Malomed$^{4}$}

\affiliation{$^1$ Institute of Spectroscopy, Russian Academy of Sciences, Troitsk, Moscow
Region, 142190, Russia\\
$^{2}$ Department of Mathematics and Statistics, University of Massachusetts,
Amherst MA 01003-4515, USA \\
$^{3}$ Department of Physics, University of Athens,
Panepistimiopolis, Zografos, Athens 157 84, Greece \\
$^{4}$ Department of Physical Electronics, Faculty of Engineering, Tel Aviv
University, Tel Aviv 69978, Israel }

\date{\today}

\begin{abstract}
Generation of wave structures by a two-dimensional object (laser beam)
moving in a two-dimensional two-component Bose-Einstein condensate with a
velocity greater than both sound velocities of the mixture is studied by
means of analytical methods and systematic simulations of the coupled
Gross-Pitaevskii equations. The wave pattern features three regions
separated by two Mach cones. Two branches of linear patterns similar to the
so-called ``ship waves'' are located outside the corresponding Mach cones,
and oblique dark solitons are found inside the wider cone. An analytical
theory is developed for the linear patterns. A particular dark-soliton
solution is also obtained, its stability is investigated, and two unstable
modes of transverse perturbations are identified. It is shown that, for a
sufficiently large flow velocity, this instability has a convective
character in the reference frame attached to the moving body, which makes
the dark soliton effectively stable. The analytical findings are
corroborated by numerical simulations.
\end{abstract}


\maketitle

\section{Introduction}

Breakdown of superfluidity at large velocities of the flow is a
fundamentally important problem which has been widely studied in physics of
quantum liquids, such as $^{4}$He and Bose-Einstein condensates (BECs) of
dilute atomic gases (see book \cite{ps2003} and references therein). As is
known, the breakdown is caused by opening of channels of radiation of
elementary excitations in the fluid, as it happens, for instance, with
increase of the velocity of a body (``obstacle") moving through the
superfluid (in the experiment, a far-blue-detuned laser beam, which produces
a local repulsive force acting on atoms, usually plays the role of the
obstacle \cite{engels07,engels08}). In the BEC, solitary waves and vortices
are readily generated if the size of the obstacle is of the order of (or
greater than) the characteristic healing length of condensate. The
generation of these structures manifests itself as an effective dissipation,
which implies the loss of superfluidity at some critical value of the
obstacle's velocity \cite{hakim,us1}. On the other hand, if the size of the
obstacle is much smaller than the healing length, the main loss channel
corresponds to the Cherenkov radiation of Bogoliubov excitations, and it
opens at supersonic velocities of the obstacle \cite{ap04}. If a large
obstacle moves at a supersonic velocity, then the amplitude of the generated
waves becomes large too. In the latter case, two dispersive shocks, which
start their propagation from the front and the rear parts of the moving
body, are formed. Far from the body, the shock front gradually transforms
into a linear ``ship wave" located outside the Mach cone \cite%
{carusotto,gegk07,gk07,gsk08}, whereas the rear zone of the shock is
converted into a ``fan" of oblique dark solitons located inside the Mach
cone \cite{ek06,egk06,egk07b}. Although, as is well known, such dark
solitons are unstable with respect to transverse perturbations, it was shown
in \cite{kp08} that for a flow velocity greater than some critical value
this instability becomes a convective one (rather than being absolute) in
the reference frame moving along with the obstacle. This fact actually means
that the dark solitons are effectively stable in the region around the
obstacle. Some of these structures have already been observed in experiments
\cite{cornell05,carusotto}, and similar non-stationary dispersive shocks
were studied both theoretically and experimentally in BEC \cite%
{damski04,kgk04,simula,hoefer,engels07} (see also review \cite{nonlin}) and
nonlinear optics \cite{fleischer,trillo,fleischer3,fleischer2,el07,khamis08}.

The picture described above corresponds to a single-component BEC whose
mean-field dynamics is governed by the respective Gross-Pitaevskii equation
\cite{ps2003}. At the same time, one of
the focal points of activity in BEC physics has been the study of
multi-component settings, a prototypical one being presented by binary
mixtures \cite{Myatt1997a,Hall1998a,Stamper-Kurn1998b}. These two-component
media exhibit phase-separation phenomena \cite{boris1,boris2,tsubota} due to
the nonlinear interaction between the different atomic species (or possibly
different hyperfine states) that constitute the mixture. The formation of
robust single- and multi-ring patterns \cite{Hall1998a,dshall}, the
evolution of initially coincident triangular vortex lattices through a
turbulent regime into an interlaced square vortex lattice~\cite%
{Schweikhard2004a} in coupled hyperfine states in the $^{87}$Rb condensate,
and the study of the interplay between atomic states at different Zeeman
levels in the $^{23}$Na condensate, forming striated magnetic domains~in
optical traps \cite{Miesner1999a,Stenger1999a}, are only a small subset
among the many possibilities that multi-component BECs can offer. It should
also be noted that mixtures with a higher number of components, namely
spinor condensates~\cite{ket1,cahn}, are known too. They have been realized
with the help of far-off-resonant optical techniques for trapping ultracold
atomic gases~\cite{ket0}, which, in turn, allowed the spin degree of freedom
to be explored (previously, it was frozen in magnetic traps).

Our aim in the present work is to unite these two areas by investigating the
effects generated by the motion of an obstacle in a ``pancake"-shaped, i.e.,
effectively two-dimensional (2D), \emph{two-component}\textit{\ }BEC. In an
earlier work \cite{us2}, the critical situation, when the obstacle had the
velocity comparable to the two speeds of sound in the two-component system,
was examined. Here, we extend the analysis to the supercritical case, when
the speed of the moving body may be significantly higher than the sound
speed(s). We demonstrate that two branches of linear, so-called
``ship-wave", patterns form outside of the two Mach cones (which are
associated with the speeds of sound), while oblique dark solitons are
located inside the wider Mach cone. While these dark solitons are unstable
at relatively low velocities of the motion of the obstacle, they can be
convectively stabilized at sufficiently high values of the velocity.

The presentation is structured as follows. In section II, we outline the
model under consideration. In section III, we study the linear (ship) waves
by means of both analytically and numerical methods. In section IV, we study
dark solitons (again, by dint of both analytical and numerical methods).
Finally, we summarize our findings in section V, and discuss potential
directions for future work.

\section{Mathematical Model and Setup}

In the usual mean-field approximation, the 2D flow of a binary condensate
mixture past an obstacle obeys the system of nonlinearly coupled
Gross-Pitaevskii equations (GPEs), with spatial coordinates $\mathbf{r}%
=(x,y) $. In the scaled form, the equations take a well-known form \cite%
{ps2003,nonlin},
\begin{equation}
\begin{split}
i\frac{\partial \psi _{1}}{\partial t}& =-\frac{1}{2}\Delta \psi
_{1}+(g_{11}\left\vert \psi _{1}\right\vert ^{2}+g_{12}\left\vert \psi
_{2}\right\vert ^{2})\psi _{1}+V(\mathbf{r},t)\psi _{1}, \\
i\frac{\partial \psi _{2}}{\partial t}& =-\frac{1}{2}\Delta \psi
_{2}+(g_{12}\left\vert \psi _{2}\right\vert ^{2}+g_{22}\left\vert \psi
_{2}\right\vert ^{2})\psi _{2}+V(\mathbf{r},t)\psi _{2}.
\end{split}
\label{1-2}
\end{equation}
We assume that atoms in both species have the same mass (i.e., they
represent different hyperfine states of the same atom, as is often the case,
see e.g., Ref. \cite{dshall} and references therein), $V(\mathbf{r},t)$
being the potential induced by the moving obstacle, which is identical for
both components.

Small-amplitude waves and solitons correspond to potential flows with zero
vorticity, for which Eqs.~(\ref{1-2}) can be transformed into a hydrodynamic
form by means of substitutions
\begin{equation}
\psi_{1}(\mathbf{r},t)=\sqrt{n_{1}(\mathbf{r},t)}\exp \left( i\int^{\br} \mathbf{u%
}_{1}(\mathbf{r}',t)d\mathbf{r}'-i\mu _{1}t\right),\quad
\psi_{2}(\mathbf{r},t)=%
\sqrt{n_{2}(\mathbf{r},t)}\exp \left( i\int^{\br} \mathbf{u}_{2}(\mathbf{r}',t)d%
\mathbf{r}'-i\mu _{2}t\right) ,  \label{1-3}
\end{equation}
where $n_{1,2}(\mathbf{r},t)$ are the atom densities of the two BEC
components, $\mathbf{u}_{1,2}(\mathbf{r},t)$ their velocity fields, and $\mu
_{1,2}$ the respective chemical potentials. Substituting Eqs.~(\ref{1-3})
into the Eqs.~(\ref{1-2}) we arrive at the following system,%
\begin{equation}
\begin{split}
n_{1,t}+\nabla \cdot (n_{1}\mathbf{u}_{1}) =0,\qquad
n_{2,t}+\nabla \cdot (n_{2}\mathbf{u}_{2})=0, \\
\mathbf{u}_{1,t}+(\mathbf{u}_{1}\cdot \nabla )\mathbf{u}_{1}+g_{11}\nabla
n_{1}+g_{12}\nabla n_{2}+\nabla \left( \frac{(\nabla n_{1})^{2}}{8n_{1}^{2}}-%
\frac{\Delta n_{1}}{4n_{1}}\right) +\nabla V(\mathbf{r},t)& =0, \\
\mathbf{u}_{2,t}+(\mathbf{u}_{2}\cdot \nabla )\mathbf{u}_{2}+g_{12}\nabla
n_{1}+g_{22}\nabla n_{2}+\nabla \left( \frac{(\nabla n_{2})^{2}}{8n_{2}^{2}}-%
\frac{\Delta n_{2}}{4n_{2}}\right) +\nabla V(\mathbf{r},t)& =0.
\end{split}
\label{1-4}
\end{equation}%
The first pair of the equations represents the conservation of the number of
atoms in each component, and the second pair corresponds to the Euler's
equations for fluid velocities under the action of the pressure induced by
interactions between atoms, the obstacle's potential, and quantum dispersion.

We consider waves generated by the obstacle moving at constant velocity $U$
along the $x$-axis,
\begin{equation}
V(\mathbf{r},t)=V(x-Ut),  \label{1-5}
\end{equation}%
through a uniform condensate, so that at $|x|\rightarrow \infty $ both
components have constant densities and vanishing velocities. This setting
implies the absence of a trapping potential in the plane of the quasi-2D BEC
(or, more realistically, a very weak trapping). In the reference frame
moving along with the obstacle, the unperturbed condensate flows at constant
velocity $\mathbf{u}=(-U,0)$, hence the respective boundary conditions for
the densities and velocities are
\begin{equation}
n_{1}\rightarrow n_{10},~n_{2}\rightarrow n_{20},~\mathbf{u}_{1}\rightarrow
\mathbf{u},~\mathbf{u}_{2}\rightarrow \mathbf{u}\quad \text{at}\quad
|x|\rightarrow \infty .  \label{1-6}
\end{equation}%
As follows from Eqs.~(\ref{1-6}), chemical potentials $\mu _{1,2}$ are
related with the asymptotic densities $n_{10},n_{20}$ by the relations
\begin{equation}
\mu _{1}=g_{11}n_{10}+g_{12}n_{20},\quad \mu _{2}=g_{12}n_{10}+g_{22}n_{20}.
\label{1-7}
\end{equation}%
In this reference frame, the wave pattern is a stationary one, which is
convenient for analytical considerations.

\section{Linear ``ship waves''}

We first aim to consider linear waves generated by the moving obstacle,
assume that potential $V(\mathbf{r},t)$ is weak enough to apply the
perturbation theory \cite{ap04} based on linearized equations. Actually, the
approximation is valid everywhere where the amplitude of the waves is small%
---in particular, far enough from the obstacle outside the Mach cones
associated with the speeds of sound. Thus, we introduce small deviations
from the uniform state, namely,
\begin{equation}
n_{1}=n_{10}+n_{1}^{\prime },\quad n_{2}=n_{20}+n_{2}^{\prime },\quad\mathbf{u}_{1}=%
\mathbf{u}+\mathbf{u}_{1}^{\prime },\quad\mathbf{u}_{2}=\mathbf{u}+\mathbf{u}%
_{2}^{\prime },  \label{2-1}
\end{equation}%
and accordingly linearize Eqs. (\ref{1-4}):

\begin{equation}
n_{1,t}^{\prime }+n_{10}(\nabla \cdot \mathbf{u}_{1}^{\prime })+(\mathbf{u}%
\cdot \nabla )n_{1}^{\prime }=0,\quad
n_{2,t}^{\prime }+n_{20}(\nabla \cdot
\mathbf{u}_{2}^{\prime })+(\mathbf{u}\cdot \nabla )n_{2}^{\prime }=0,
\end{equation}
\begin{equation}
\begin{split}
\mathbf{u}_{1,t}^{\prime }+(\mathbf{u}\cdot \nabla )\mathbf{u}_{1}^{\prime
}+g_{11}\nabla n_{1}^{\prime }+g_{12}\nabla n_{2}^{\prime }-\frac{1}{4n_{10}}%
\nabla (\Delta n_{1}^{\prime })& =-\nabla V, \\
\mathbf{u}_{2,t}^{\prime }+(\mathbf{u}\cdot \nabla )\mathbf{u}_{2}^{\prime
}+g_{12}\nabla n_{1}^{\prime }+g_{22}\nabla n_{2}^{\prime }-\frac{1}{4n_{20}}%
\nabla (\Delta n_{2}^{\prime })& =-\nabla V.
\end{split}
\label{2-2}
\end{equation}%
In the stationary case, all time derivatives vanish. Further, we apply the
Fourier transform,
\begin{equation}
n_{1}^{\prime }(\mathbf{r},t)=\int \int \tilde{n}_{1}^{\prime }(\mathbf{k}%
,t)e^{i\mathbf{kr}}\frac{d^{2}k}{(2\pi )^{2}}.  \label{2-2a}
\end{equation}%
Then, we eliminate the velocity perturbations in Eqs.~(\ref{2-2}) in favor
of the densities, arriving at the following system of two linear equations:
\begin{equation}
\begin{split}
& [-(\mathbf{k}\cdot \mathbf{u})^{2}+k^{2}(g_{11}n_{10}+k^{2}/4)]\tilde{n}%
_{1}^{\prime }+k^{2}g_{12}n_{10}\tilde{n}_{2}^{\prime }=-k^{2}\tilde{V}%
n_{10}, \\
& k^{2}g_{12}n_{20}\tilde{n}_{1}^{\prime }+[-(\mathbf{k}\cdot \mathbf{u}%
)^{2}+k^{2}(g_{22}n_{20}+k^{2}/4)]\tilde{n}_{2}^{\prime }=-k^{2}\tilde{V}%
n_{20},
\end{split}
\label{2-3}
\end{equation}%
where tildes denote the Fourier components. This linear system can be
readily solved for $\tilde{n}_{1,2}^{\prime }$ and the inverse
Fourier transform yields
\begin{equation}
\begin{split}
n_{1}^{\prime }& =-n_{10}\int \int \frac{k^{2}\tilde{V}\left\{ [-(\mathbf{k}%
\cdot \mathbf{u})^{2}+k^{2}(g_{22}n_{20}+k^{2}/4)]-k^{2}g_{12}n_{20}\right\}
}{((\mathbf{k}\cdot \mathbf{u})^{2}-\omega _{+}^{2})((\mathbf{k}\cdot
\mathbf{u})^{2}-\omega _{-}^{2})}\frac{d^{2}k}{(2\pi )^{2}}, \\
n_{2}^{\prime }& =-n_{20}\int \int \frac{k^{2}\tilde{V}\left\{ [-(\mathbf{k}%
\cdot \mathbf{u})^{2}+k^{2}(g_{11}n_{10}+k^{2}/4)]-k^{2}g_{12}n_{10}\right\}
}{((\mathbf{k}\cdot \mathbf{u})^{2}-\omega _{+}^{2})((\mathbf{k}\cdot
\mathbf{u})^{2}-\omega _{-}^{2})}\frac{d^{2}k}{(2\pi )^{2}},
\end{split}
\label{2-4}
\end{equation}%
where the dispersion relations for the linear waves in the binary mixture
are given by
\begin{equation}
\omega _{\pm }^{2}=\tfrac{1}{2}k^{2}\left[ g_{11}n_{10}+g_{22}n_{20}+\tfrac{1%
}{2}k^{2}\pm \sqrt{(g_{11}n_{10}-g_{22}n_{20})^{2}+4g_{12}^{2}n_{10}n_{20}}%
\right] ,  \label{2-5}
\end{equation}%
In the long-wave limit, Eqs.~(\ref{2-5}) yield the sound velocities \cite%
{us2}, $c_{\pm }=\omega (k)/k$, given by:%
\begin{equation}
c_{\pm }\equiv \lim_{k\to0} \frac{\omega (k)}{k}=\sqrt{\frac{%
g_{11}n_{10}+g_{22}n_{20}\pm \sqrt{%
(g_{11}n_{10}-g_{22}n_{20})^{2}+4g_{12}^{2}n_{10}n_{20}}}{2}}.  \label{2-6}
\end{equation}

Expressions (\ref{2-4}) can be analyzed following the lines of Ref.~\cite%
{gk07,gsk08}. To this end, we introduce new coordinates (see Fig.~1),
\begin{equation}
\mathbf{r}\equiv (-r\cos \chi ,r\sin \chi ),\quad \mathbf{k}\equiv (k\cos
\eta ,k\sin \eta ),  \label{2-7}
\end{equation}%
and assume that the wavelength of the pattern is much greater than the
characteristic size of the obstacle, hence its potential can be approximated
by the form $V(\mathbf{r})=V_{0}\delta (\mathbf{r})$. Then, we obtain
\begin{equation}
\begin{split}
n_{1}^{\prime }& =\frac{4V_{0}n_{10}}{\pi ^{2}}\int_{-\pi }^{\pi
}\int_{0}^{\infty }\frac{k[(g_{22}-g_{12})n_{20}+k^{2}/4-U^{2}\cos ^{2}\eta
]e^{i\mathbf{k}\mathbf{r}}}{(k^{2}-k_{+}^{2}-i0)(k^{2}-k_{-}^{2}-i0)}dkd\eta
, \\
n_{2}^{\prime }& =\frac{4V_{0}n_{20}}{\pi ^{2}}\int_{-\pi }^{\pi
}\int_{0}^{\infty }\frac{k[(g_{11}-g_{12})n_{10}+k^{2}/4-U^{2}\cos ^{2}\eta
]e^{i\mathbf{k}\mathbf{r}}}{(k^{2}-k_{+}^{2}-i0)(k^{2}-k_{-}^{2}-i0)}dkd\eta
,
\end{split}
\label{2-8}
\end{equation}%
where infinitesimal imaginary parts in the denominators are written to
define correct contributions from the poles corresponding to adiabatically
slow switching on of the potential, and
\begin{equation}
k_{\pm }\equiv 2\sqrt{U^{2}\cos ^{2}\eta -c_{\pm }^{2}}.  \label{2-9}
\end{equation}

\begin{figure}[tb]
\begin{center}
\includegraphics[width=6cm,height=6cm,clip]{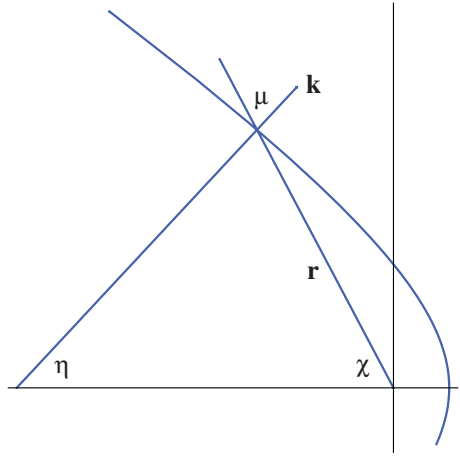}
\end{center}
\caption{(Color online.) Coordinates defining radius-vector $\mathbf{r}$ and
wave vector $\mathbf{k}$. The latter one is normal to the wave front of one
of the ship-wave modes, which is shown schematically by a curve.}
\label{fig1}
\end{figure}

Next, we split the integration domain of $\eta $ into two parts: $\int_{-\pi
/2}^{3\pi /2}d\eta \equiv \int_{-\pi /2}^{\pi /2}d\eta +\int_{\pi /2}^{3\pi
/2}d\eta $. After replacement $\eta ^{\prime }=\eta -\pi $ in the second
term, one can notice that the integrand turns into its own complex
conjugate, allowing one to write the integrals in Eq.~(\ref{2-8}) as
\begin{equation}
\begin{split}
n_{1}^{\prime }& =\frac{8V_{0}n_{10}}{\pi ^{2}}\mathrm{Re}\int_{-\pi
/2}^{\pi /2}d\eta \int_{0}^{\infty }\frac{%
k[(g_{22}-g_{12})n_{20}+k^{2}/4-U^{2}\cos ^{2}\eta ]e^{i\mathbf{k}\mathbf{r}}%
}{(k^{2}-k_{+}^{2}-i0)(k^{2}-k_{-}^{2}-i0)}dk, \\
n_{2}^{\prime }& =\frac{8V_{0}n_{20}}{\pi ^{2}}\mathrm{Re}\int_{-\pi
/2}^{\pi /2}d\eta \int_{0}^{\infty }\frac{%
k[(g_{11}-g_{12})n_{10}+k^{2}/4-U^{2}\cos ^{2}\eta ]e^{i\mathbf{k}\mathbf{r}}%
}{(k^{2}-k_{+}^{2}-i0)(k^{2}-k_{-}^{2}-i0)}dk.
\end{split}
\label{2-10}
\end{equation}%
The integration over $k$ should be carried out over the positive real
half-axis. However, we can add to this path a quarter of an infinite circle
and the imaginary half-axis in the complex $k$ plane, to build a closed
integration contour. It is easy to show that the contribution from the
quarter of the infinite circle is zero, and contribution from the imaginary
axis depends on $r$ as $r^{-2}$, decaying at large $r$ much faster than the
contribution of the pole, which is $\sim r^{-1/2}$. Thus, far from the
obstacle, it is sufficient to keep only the contribution from the poles,
which yields ($\nu \equiv \pi -\chi -\eta $)
\begin{equation}
\begin{split}
n_{1}^{\prime }& =-\frac{2V_{0}n_{10}}{\pi (c_{+}^{2}-c_{-}^{2})}\left\{
[c_{+}^{2}-(g_{22}-g_{12})n_{20}]\mathrm{Im}\int_{-\pi /2}^{\pi /2}d\eta
e^{ik_{+}r\cos \nu }-[c_{-}^{2}-(g_{22}-g_{12})n_{20}]\mathrm{Im}\int_{-\pi
/2}^{\pi /2}d\eta e^{ik_{-}r\cos \nu }\right\} , \\
n_{2}^{\prime }& =-\frac{2V_{0}n_{20}}{\pi (c_{+}^{2}-c_{-}^{2})}\left\{
[c_{+}^{2}-(g_{11}-g_{12})n_{10}]\mathrm{Im}\int_{-\pi /2}^{\pi /2}d\eta
e^{ik_{+}r\cos \nu }-[c_{-}^{2}-(g_{11}-g_{12})n_{10}]\mathrm{Im}\int_{-\pi
/2}^{\pi /2}d\eta e^{ik_{-}r\cos \nu }\right\} .
\end{split}
\label{2-11}
\end{equation}

Far from the obstacle, where phases $\mathbf{k}_{\pm }\mathbf{r}=rs_{\pm }$
are large, we are dealing with large values of
\begin{equation}
s_{\pm }(\eta )=k_{\pm }(\eta )\cos (\chi +\eta ),  \label{2-12}
\end{equation}%
and the integrals in Eq.~(\ref{2-11}) can be estimated by means of the
standard stationary-phase method. Since calculations of both integrals are
identical, we consider, for definiteness, the integral over $k_{+}$.
Condition $\partial s_{+}/\partial \eta =0$ is an equation for the
stationary-phase point, which can be easily transformed to
\begin{equation}
\tan \nu _{+}=\left( 2U^{2}/k_{+}^{2}\right) \sin 2\eta _{+},  \label{2-13}
\end{equation}%
or, with regard to the definition of $\nu $, we obtain an expression for $%
\chi $,
\begin{equation}
\tan \chi _{+}=\frac{(1+k_{+}^{2}/(2c_{+}^{2}))\tan \eta _{+}}{%
U^{2}/c_{+}^{2}-(1+k_{+}^{2}/(2c_{+}^{2}))}.  \label{2-14}
\end{equation}%
Here $\eta _{+}$ takes values in interval
\begin{equation}
-\arccos \frac{1}{M_{+}}\leq \eta _{+}\leq \arccos \frac{1}{M_{+}},\quad
M_{+}\equiv \frac{U}{c_{+}},  \label{2-15}
\end{equation}%
while the corresponding vector $\left\{ x,y\right\} $, given by the
parametric expressions,
\begin{equation}
x(\eta _{+})=\frac{4\phi c_{+}^{2}}{k_{+}^{3}}(M_{+}^{2}\cos 2\eta
_{+}-1)\cos \eta _{+},\quad y(\eta _{+})=\frac{4\phi c_{+}^{2}}{k_{+}^{3}}%
(2M_{+}^{2}\cos ^{2}\eta _{+}-1)\sin \eta _{+},  \label{2-16}
\end{equation}%
moves along a curve with constant phase $\phi $ (e.g., a crest line).

As usual in the method of stationary phase, we reduce the integrals in
(\ref{2-11}) to Gaussian ones and, as a final result, obtain
\begin{equation}\label{2-17}
\begin{split}
n_{1}^{\prime }=&-\frac{2V_{0}n_{10}}{\pi (c_{+}^{2}-c_{-}^{2})}\Bigg\{\left[
c_{+}^{2}-(g_{22}-g_{12})n_{20}\right] \sqrt{\frac{2\pi }{k_{+}r}}\\
&\times \frac{\left[ 1+(4U^{2}/k_{+}^{2})\sin ^{2}\left( 2\eta _{+}\right) %
\right] ^{1/4}}{\left[ 1+\left( 4U^{2}/k_{+}^{2}\right) \cos 2\eta
_{+}+\left( 12U^{4}/k_{+}^{4}\right) \sin ^{2}\left( 2\eta _{+}\right) %
\right] ^{1/2}}\cos (k_{+}r\cos \nu _{+}-\pi /4)\\
&-\left[ c_{-}^{2}-(g_{22}-g_{12})n_{20}\right] \sqrt{\frac{2\pi }{k_{-}r}}%
\frac{\left( 1+\left( 4U^{2}/k_{-}^{2}\right) \sin ^{2}2\eta _{-}\right)
^{1/4}}{\left[ 1+\left( 4U^{2}/k_{-}^{2}\right) \cos \left( 2\eta
_{-}\right) +\left( 12U^{4}/k_{-}^{4}\right) \sin ^{2}\left( 2\eta
_{-}\right) \right] ^{1/2}}\\
&\times \cos (k_{-}r\cos \nu _{-}-\pi /4)\Bigg\}
\end{split}
\end{equation}%
where $\nu _{\pm }\equiv \pi -\chi _{\pm }-\eta _{\pm }$, and similar
expressions can be derived for the density oscillations of the second
component of the binary condensate.

An example of wave patterns generated by numerical simulations of
Eqs.~(\ref{1-2}) is displayed in Fig.~2, for parameters $g_{11}=1.5,%
\,g_{22}=1.25,\,g_{12}=1.0,$ and $n_{10}=1.0,\,n_{20}=2.0$ and velocity $%
U=3.5$.
\begin{figure}[tb]
\subfigure{
         \begin{minipage}[b]{0.45\textwidth}
         \centering
         \includegraphics[scale=0.40]{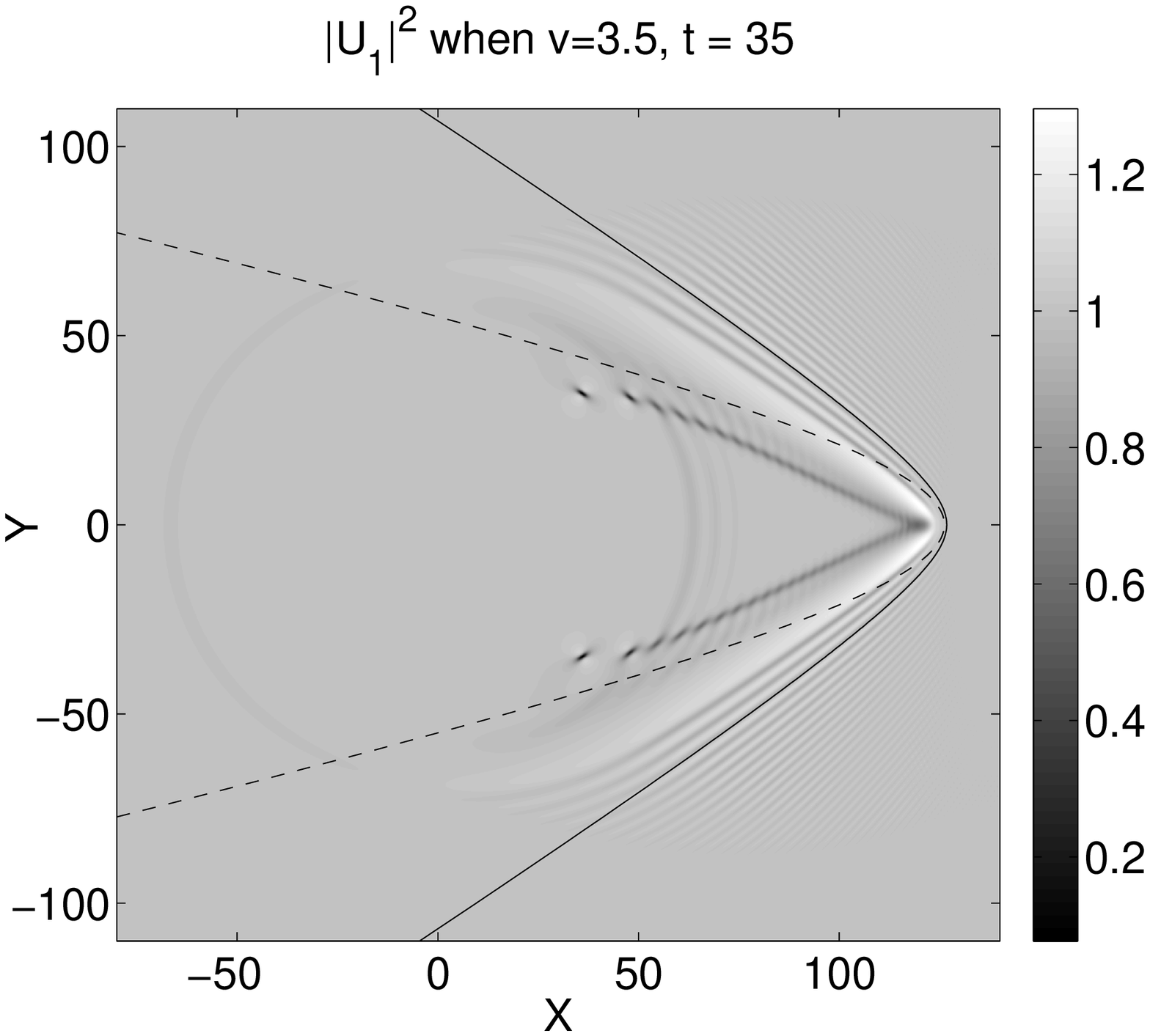}
         \end{minipage}}
\subfigure{
         \begin{minipage}[b]{0.45\textwidth}
         \centering
         \includegraphics[scale=0.40]{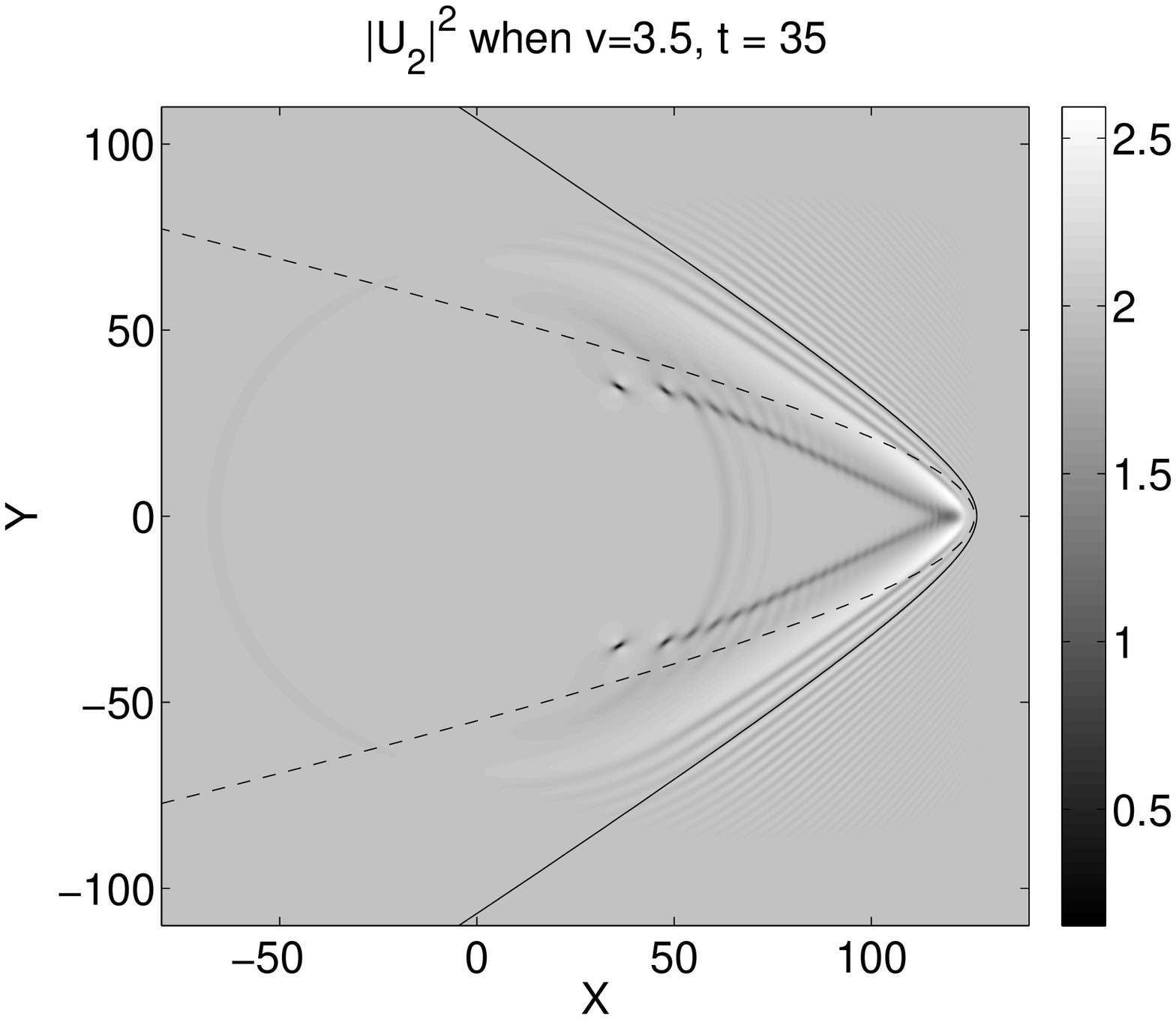}
         \end{minipage}}
\caption{Spatial contour plots of densities of the two components (left and
right panels, respectively) of the BEC for the flow velocity $U=3.5$, at $%
t=35$. The form of the obstacle'e potential used in this case is $V(x,y,t)=2%
\left[ \text{sech}\left( \protect\sqrt{(x-Ut)^{2}+y^{2}}/2\right) \right]
^{2}$. Solid and dashed thin lines correspond to wave crests of two modes
of ship waves.}
\label{fig2}
\end{figure}
Three different types of waves can clearly be distinguished in this picture.
First, one observes ship waves outside the Mach cones, which were analyzed
above. Oblique dark solitons, which are located inside the outer Mach cone,
they will be discussed in the next Section. One can also note that the
oblique solitons are modulated by concentric circular waves, which were
actually generated by the initial introduction (switching on) of the
obstacle's potential. These circular waves are not related to the steady
regime and will not be considered here.

We have compared analytical results (\ref{2-16}) for the crest lines with
the numerical findings. The analytically predicted curves are shown by thin
solid and dashed lines in Fig.~2. To simplify the pattern, we have chosen
the parameters of the binary BEC so that the chemical potentials (\ref{1-7})
of both components are equal to each other. As a result, the smaller sound velocity
satisfies the relation $c_-^2=(g_{11}-g_{12})n_{10}=(g_{22}-g_{12})n_{20}$
and hence the last term in Eq.~(\ref{2-17}) as well as in analogous expression
for $n_2'$ vanishes. Therefore the ship waves in the region
between the two Mach cones are not visible in agreement with the results of
numerical simulations.
Another linear mode describes the ship waves
outside of the outer Mach cone, and the analytically calculated form of the
crest line demonstrates excellent agreement with
the numerically obtained one.

\section{Oblique dark solitons}

As mentioned above, in Fig.~2 one can observe oblique dark
solitons located inside the outer Mach cone. It is worthy to note
that they decay into vortices at end points, which is a result of
the ``snaking" instability of dark soliton stripes in 2D
\cite{kp-1970,zakharov-1975,kt-1988} (see also
\cite{mplb}). However, for large velocities $U$ this instability is
convective only \cite{kp08}, which means that the dark solitons
are effectively stable around the obstacle and, hence, the length
of the dark solitons increases with time. The solitons originate
from a depression in the density distribution formed behind the
obstacle by the flow, therefore their depth also varies near the
obstacle. However, upon sufficiently long evolution time, there
exists a region where the oblique dark solitons can be considered
as quasi-stationary structures. In this region, the solitons are
described by the stationary solution of the
GPEs. This allows us to take the stationary GPEs in the hydrodynamic form (%
\ref{1-4}),
\begin{equation}
\begin{split}
\nabla \cdot (n_{1}\mathbf{u}_{1})=0,\quad \nabla \cdot (n_{2}\mathbf{u}_{2})=0,\\
(\mathbf{u}_{1}\cdot \nabla )\mathbf{u}_{1}+g_{11}\nabla n_{1}+g_{12}\nabla
n_{2}+\nabla \left( \frac{(\nabla n_{1})^{2}}{8n_{1}^{2}}-\frac{\Delta n_{1}%
}{4n_{1}}\right) & =0, \\
(\mathbf{u}_{2}\cdot \nabla )\mathbf{u}_{2}+g_{12}\nabla n_{1}+g_{22}\nabla
n_{2}+\nabla \left( \frac{(\nabla n_{2})^{2}}{8n_{2}^{2}}-\frac{\Delta n_{2}%
}{4n_{2}}\right) & =0.
\end{split}
\label{3-1}
\end{equation}%
Here, it is also taken into regard that the obstacle's potential is
negligible far from it. Equations (\ref{3-1}) should be solved with the
boundary conditions
\begin{equation}
n_{1}\rightarrow n_{10},\quad n_{2}\rightarrow n_{20},~\mathbf{u}%
_{1}\rightarrow (-U,0),~\mathbf{u}_{2}\rightarrow (-U,0)\quad \text{at}\quad
|x|\rightarrow \infty ,  \label{3-2}
\end{equation}%
where $-U$ is the common velocity of both components relative to the
obstacle. Under the assumption that the solution depends only on
\begin{equation}
\xi =\frac{x-ay}{\sqrt{1+a^{2}}},  \label{xi}
\end{equation}%
where $a$ determines the slope of the oblique dark soliton, this system can
be readily reduced to equations
\begin{equation}
\begin{split}
\tfrac{1}{8}(n_{1,\xi }^{2}-2n_{1}n_{1,\xi \xi
})+g_{11}n_{1}^{3}+g_{12}n_{1}^{2}n_{2}+\tfrac{1}{2}qn_{10}^{2}-(\tfrac{1}{2}%
q+\mu _{1})n_{1}^{2}& =0, \\
\tfrac{1}{8}(n_{2,\xi }^{2}-2n_{2}n_{2,\xi \xi
})+g_{12}n_{1}n_{2}^{2}+g_{22}n_{2}^{3}+\tfrac{1}{2}qn_{20}^{2}-(\tfrac{1}{2}%
q+\mu _{2})n_{2}^{2}& =0,
\end{split}
\label{3-3}
\end{equation}%
where $\mu _{1}$ and $\mu _{2}$ %
are the chemical potentials defined above in Eq.~(\ref{1-7}), and
\begin{equation}
q\equiv \frac{U^{2}}{1+a^{2}}.  \label{3-5}
\end{equation}%
The flow velocities are related to the densities as follows:
\begin{equation}
{\mathbf{u}}_{i}=\left( \frac{(n_{i0}+a^{2}n_{i})U}{(1+a^{2})n_{i}},-\frac{%
aU(n_{i0}-n_{i})}{(1+a^{2})n_{i}}\right) ,\quad i=1,2.  \label{3-5a}
\end{equation}

In general, system (\ref{3-3}) has to be solved numerically. However, if the
chemical potentials of the two components are equal,
\begin{equation}
\mu _{1}=\mu _{2}=\mu ,  \label{3-6}
\end{equation}%
the system admits a simple analytical solution in a closed form. In this
case, we look for the solution as $n_{1}=n_{10}f(\xi ),$ $n_{2}=n_{20}f(\xi
) $, reducing both equations (\ref{3-3}) to a single one,
\begin{equation}
\tfrac{1}{8}(f_{\xi }^{2}-2ff_{\xi \xi })+\mu f^{3}+\tfrac{1}{2}q-(\tfrac{1}{%
2}q+\mu )f^{2}=0.  \label{3-7}
\end{equation}%
Dark-soliton solutions to Eq.~(\ref{3-7}) are known \cite{egk06}:
\begin{equation}
n_{1}=n_{1s}=n_{10}f(\xi ),\quad n_{2}=n_{2s}=n_{20}f(\xi ),\quad f(\xi )=1-%
\frac{1-q/c_{+}^{2}}{\cosh ^{2}\left[ \sqrt{c_{+}^{2}-q}\,(x-ay)/\sqrt{%
1+a^{2}}\right] },  \label{3-8}
\end{equation}%
where we have also taken into regard that condition (\ref{3-6}) leads to the
following expressions for the sound velocities (for definiteness, we
suppose here that $g_{11},g_{22}>g_{12}$, i.e., the inter-species repulsion
is weaker than the repulsive self-interactions of the two components):
\begin{equation}
c_{-}^{2}=(g_{11}-g_{12})n_{10},\quad c_{+}^{2}=\mu .  \label{3-9}
\end{equation}

\begin{figure}[tb]
\subfigure{
         \begin{minipage}[b]{0.45\textwidth}
         \centering
         \includegraphics[scale=0.40]{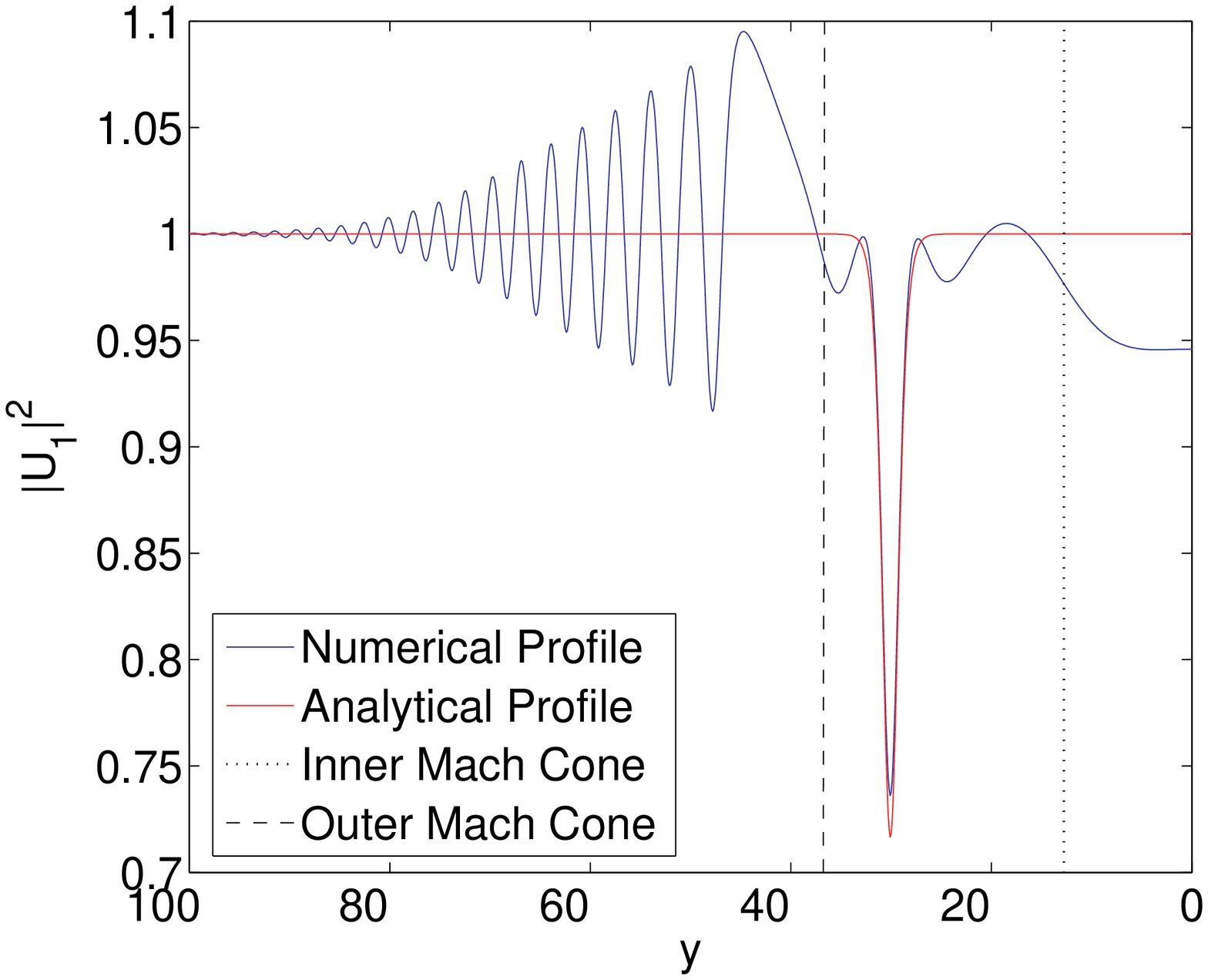}
         \end{minipage}}
\subfigure{
         \begin{minipage}[b]{0.45\textwidth}
         \centering
         \includegraphics[scale=0.40]{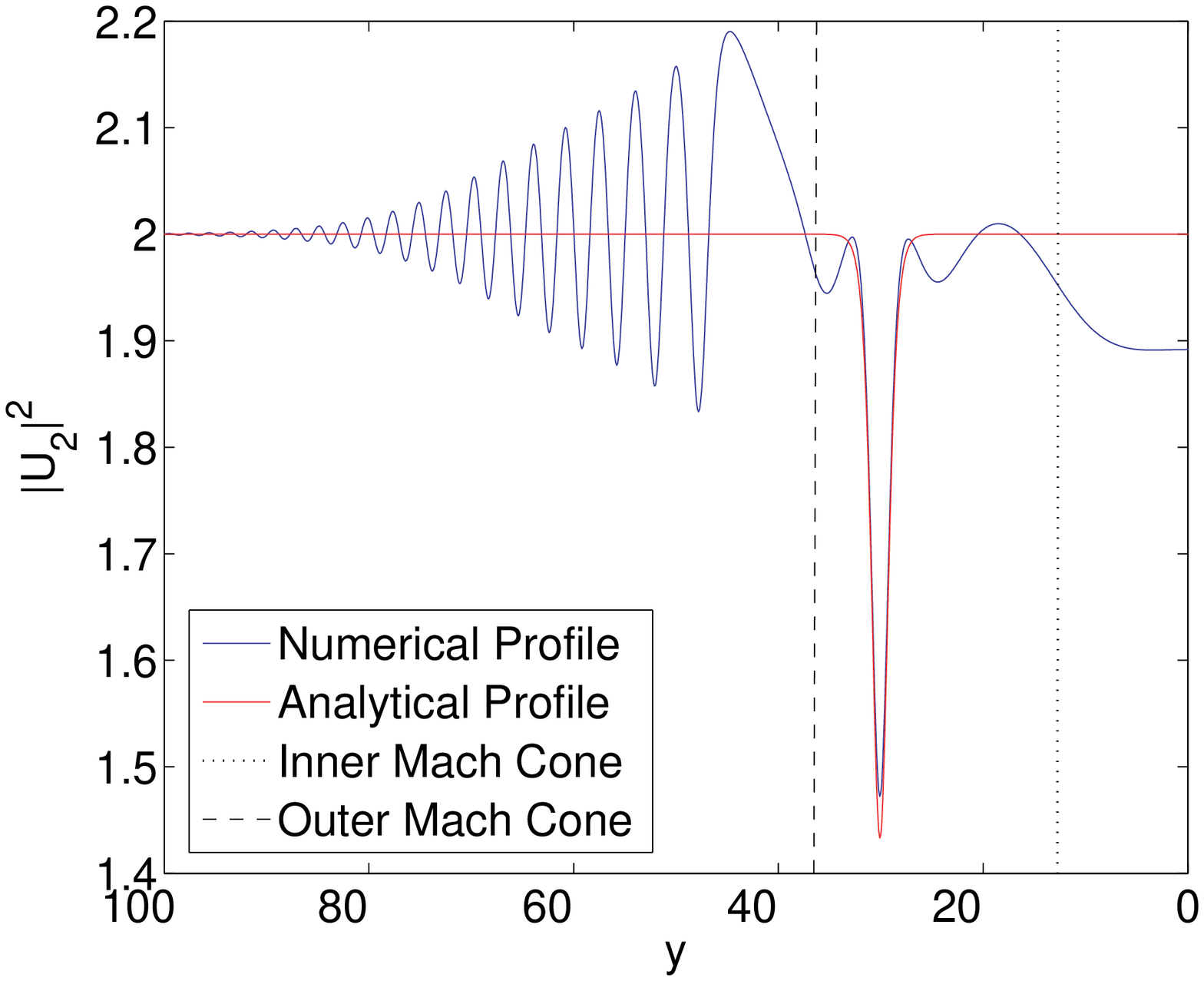}
         \end{minipage}}
\subfigure{
         \begin{minipage}[b]{0.45\textwidth}
         \centering
         \includegraphics[scale=0.40]{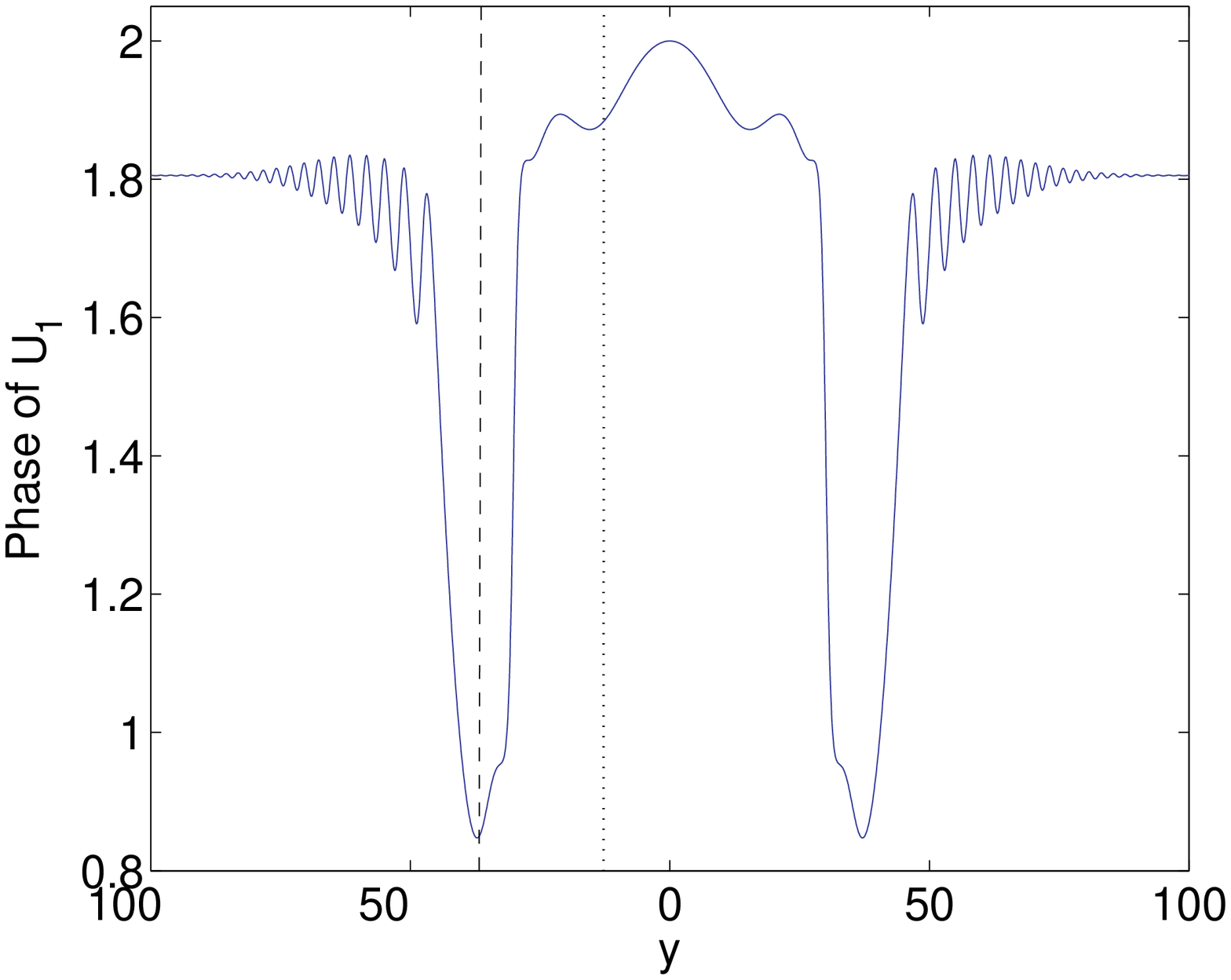}
         \end{minipage}}
\subfigure{
         \begin{minipage}[b]{0.45\textwidth}
         \centering
         \includegraphics[scale=0.40]{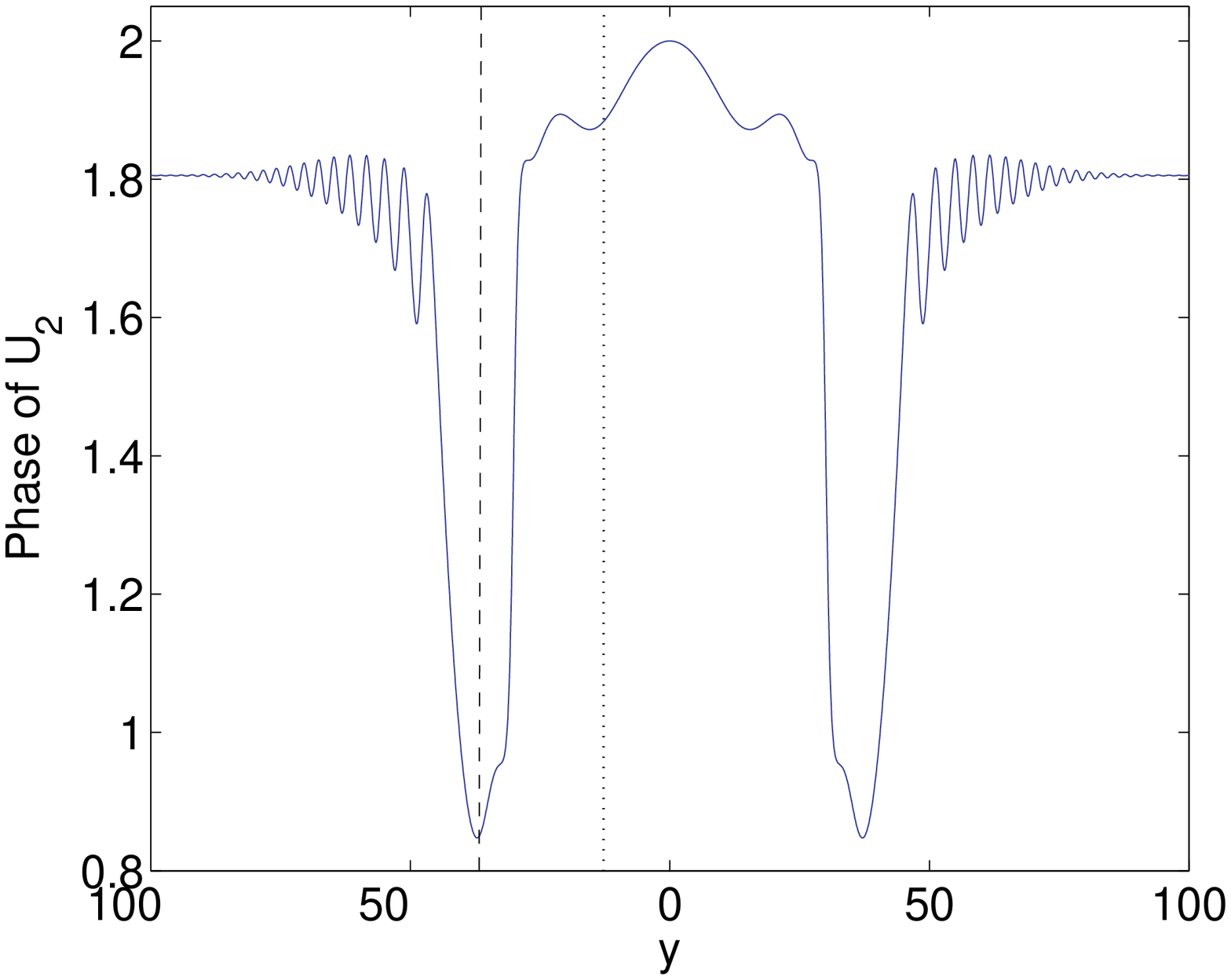}
         \end{minipage}}
\caption{(Color online.) Plots of densities (top) and phases (bottom) of the
two BEC components for flow velocity $U=4.5$. The positions of the Mach
cones are shown by dashed lines, making it evident that the oblique dark
solitons are located inside the outer Mach cone, in agreement with Eq. (%
\protect\ref{3-12}). The chemical potentials are $\protect\mu _{1}=\protect%
\mu _{2}=3.5$, cf. Eq.~(\protect\ref{3-6}), and the configurations are shown
at $t=30$.}
\label{fig3}
\end{figure}

Obviously, solution (\ref{3-8}) exists if the condition $q<c_{+}^{2}$ is
satisfied. If we introduce angle $\theta $ between the direction of the flow
and the orientation of the dark-soliton stripe, so that $a=\cot \theta $,
the latter condition can be transformed into
\begin{equation}
\sin ^{2}\theta <\frac{c_{+}^{2}}{U^{2}}=\frac{1}{M_{+}^{2}},\quad
M_{+}\equiv \frac{U}{c_{+}}.  \label{3-12}
\end{equation}%
Thus, the soliton must be located inside the outer Mach cone, which is
defined by the equation
\begin{equation}
\sin \theta _{+}=\frac{1}{M_{+}}.  \label{3-13}
\end{equation}
Although we have arrived at this conclusion under assumption (\ref{3-6}), we
conjecture that it is correct too in the general case of unequal chemical
potentials, which is confirmed by the fact that our numerical simulations
always produced oblique solitons confined inside the outer Mach cone. The
respective numerically generated profiles of the densities are shown in Fig.~%
\ref{fig3} as a function of $y$ at a fixed value of $x$; the corresponding
phase profiles are also shown in the figure. It is observed that the oblique
solitons are indeed located inside the outer Mach cone [as defined by Eq.~(%
\ref{3-13})]) but outside of the inner cone, which is defined by $\sin
\theta _{-}=1/M_{-}$. Profiles of the solitons' densities are close to the
analytically predicted ones and the jumps of phases are also in agreement
with the expected dark-soliton behavior.

\begin{figure}[tb]
\subfigure{
         \begin{minipage}[b]{0.45\textwidth}
         \centering
         \includegraphics[scale=0.40]{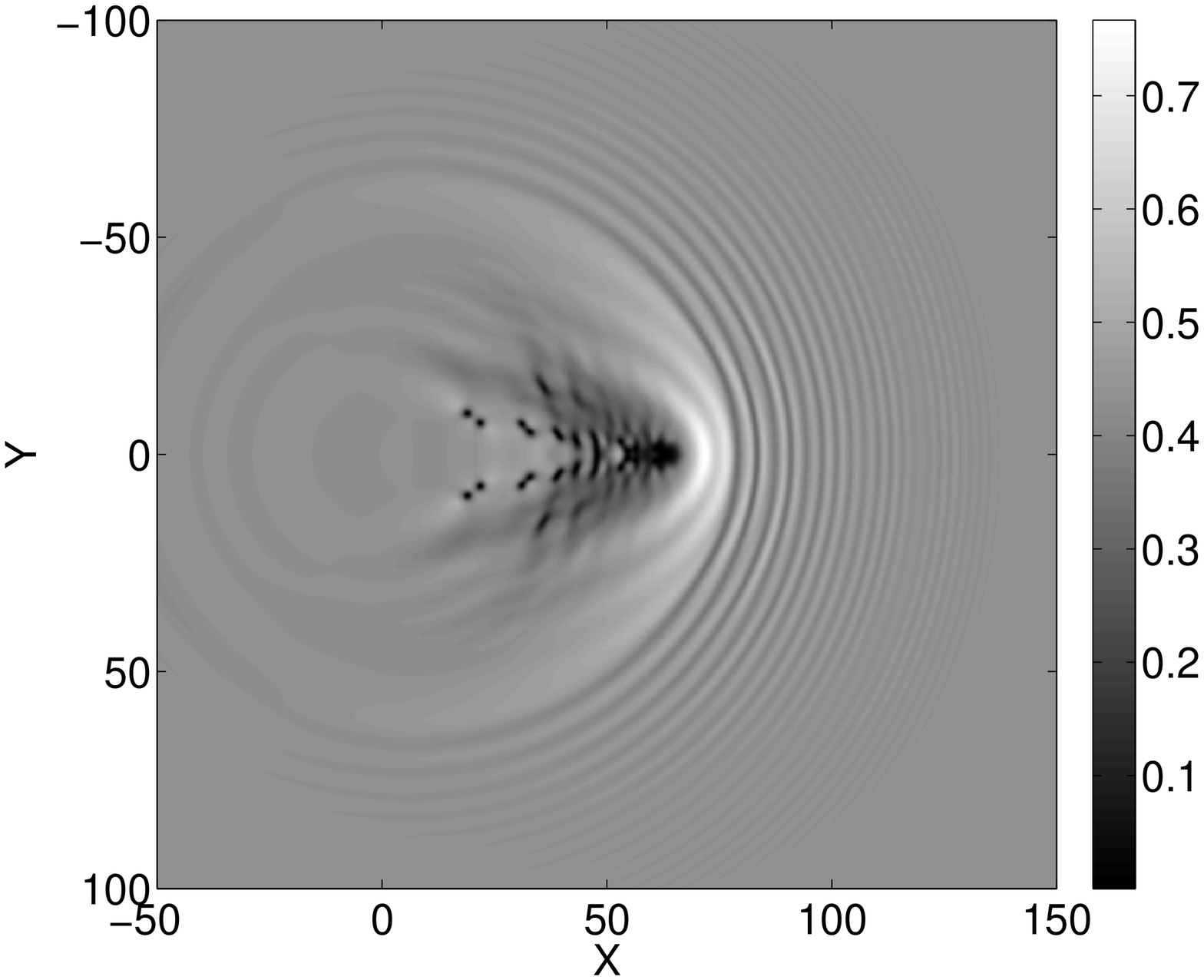}
         \end{minipage}}
\subfigure{
         \begin{minipage}[b]{0.45\textwidth}
         \centering
         \includegraphics[scale=0.40]{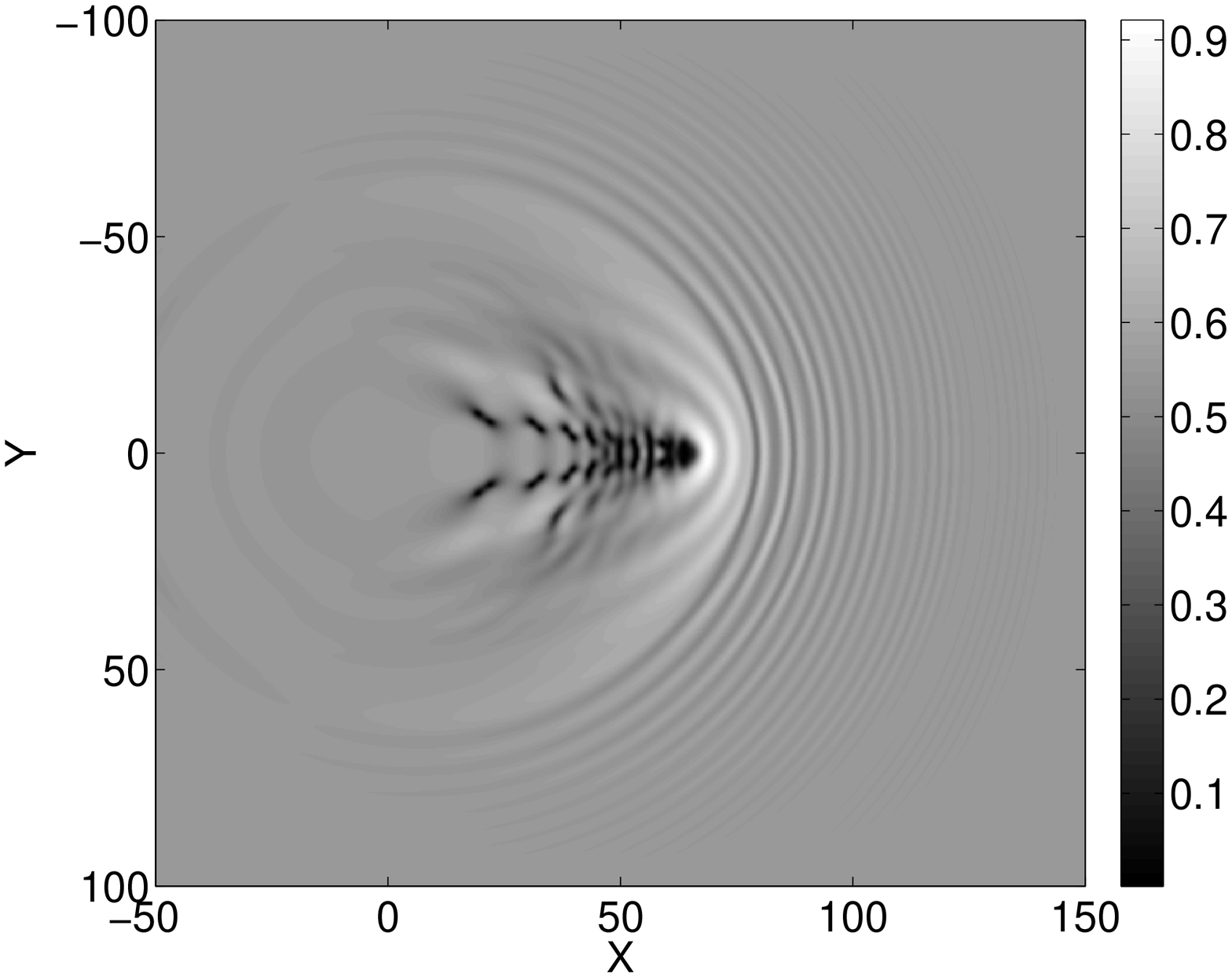}
         \end{minipage}}
\caption{Plots of densities of two BEC components for flow velocity $U=1.2$
at $t=55$.}
\label{fig4}
\end{figure}

According to condition (\ref{3-12}), solution (\ref{3-8}) exists for any
supersonic flow with $U>c_{+}$, the same being true for the existence of
numerical solutions to system (\ref{3-3}) in the general case, $\mu _{1}\neq
\mu _{2}$. However, that does not mean that the oblique dark solitons can be
generated by any such flow. While the numerical results presented in Fig.~%
\ref{fig2} show that oblique solitons indeed exist for velocity $U=3.5$, the
opposite situation, when oblique dark solitons do not emerge, is presented
in Fig.~\ref{fig4} by means of density patterns which correspond to a lower
supersonic flow velocity, $U=1.2$.
It is seen that vortex streets are generated in the latter case, rather than
dark solitons. Such a behavior is related to the well-known instability of
dark solitons with respect to transverse perturbations \cite%
{kp-1970,zakharov-1975,kt-1988}. The instability splits dark solitons into
vortex-antivortex pairs, hence dark solitons cannot develop from the density
depression behind the obstacle moving at a relatively low velocity. However,
the numerical simulations presented in Fig.~\ref{fig2} indicate that the
oblique solitons become effectively stable if the flow velocity is
sufficiently high, as first was noticed in Ref.~\cite{egk06} for the case of
a one-component BEC. The stabilization was explained in Ref.~\cite{kp08} as
a transition from the absolute instability of dark solitons to their
convective instability, at some critical value of the flow velocity, $U_{%
\mathrm{cr}}\geq c_{+}$, so that the unstable disturbances are carried away
by the flow from the region around the obstacle where, as a result, the dark
solitons look as effectively stable objects. Here, we aim to consider such a
stabilization transition for the case of the two-component BEC, which
features two unstable modes of perturbations around the dark soliton.

To find the spectrum of small-amplitude linear waves propagating along the
soliton, we now consider this solution in the reference frame in which the
condensate has zero velocity far from the obstacle. To this end, we rotate
the coordinate system by angle $\varphi =\arctan a$ [recall $a$ determines
the orientation of the dark soliton, according to Eq.~(\ref{xi})], and
perform the Galilean transformation to the frame moving relative to the
obstacle at velocity $(U\cos \varphi ,U\sin \varphi )$:
\begin{equation}
\begin{split}
\widetilde{x}& =x\cos \varphi -y\sin \varphi -U\cos \varphi \cdot t, \\
\widetilde{y}& =x\sin \varphi +y\cos \varphi -U\sin \varphi \cdot t.
\end{split}
\label{4-1}
\end{equation}%
After the transformation, velocity fields (\ref{3-5a}) become
\begin{equation}
\widetilde{\mathbf{u}}_{1}=\left( v(n_{10}/n_{1s}-1),0\right) ,\quad \widetilde{%
\mathbf{u}}_{2}=\left( v(n_{20}/n_{2s}-1),0\right) ,  \label{4-2}
\end{equation}%
and the densities take the form of
\begin{equation}
\begin{split}
\widetilde{n}_{1s}& =n_{10}f(\zeta ),\quad \widetilde{n}_{2s}=n_{20}f(\zeta ), \\
f(\zeta )& =1-\frac{1-v^{2}/c_+^2}{\cosh ^{2}\left[ \sqrt{c_{+}^{2}-v^{2}}\zeta %
\right] },\quad\zeta =\widetilde{x}-vt,
\end{split}
\label{4-3}
\end{equation}%
where the soliton's velocity in the new reference frame is
\begin{equation}
v=\frac{U}{\sqrt{1+a^{2}}}.  \label{4-4}
\end{equation}%
Below, we omit tildes attached to the new variables.

We take small transverse perturbations of the dark-soliton solution as
\begin{equation}
\begin{split}
\psi _{1}& =\psi _{1s}(\zeta )+(\psi _{1}^{\prime }+i\psi _{1}^{\prime
\prime })\exp \left( i\phi _{1s}(\zeta )-{i\mu _{1}t}\right) , \\
\psi _{2}& =\psi _{2s}(\zeta )+(\psi _{1}^{\prime }+i\psi _{2}^{\prime
\prime })\exp \left( i\phi _{2s}(\zeta )-{i\mu _{2}t}\right) ,
\end{split}
\label{4-5}
\end{equation}%
where the unperturbed solution depends only on $\zeta =x-vt$,
\begin{equation}
\psi _{js}=\sqrt{{n}_{js}}\exp \left( i\phi _{js}(\zeta )-{i\mu _{j}t}%
\right) ,  \label{4-5a}
\end{equation}%
and phases $\phi _{js}$ are related to the densities by equations
\begin{equation}
\frac{\partial \phi _{js}}{\partial \zeta }=v\left( \frac{n_{j0}}{n_{js}}%
-1\right) .  \label{4-5b}
\end{equation}%
Perturbations $\psi ^{\prime }$ and $i\psi ^{\prime \prime }$ depend on $y$
and $t$ as $\exp (ipy+\Gamma t)$. Substitution of expressions (\ref{4-5})
into Eqs.~(\ref{1-2}) and the linearization with respect to $\psi ^{\prime }$
and $i\psi ^{\prime \prime }$ lead to an eigenvalue problem,
\begin{equation}
\left(
\begin{array}{cccc}
A_{1} & -L_{I1} & 0 & 0 \\
L_{R1} & A_{1} & B & 0 \\
0 & 0 & A_{2} & -L_{I2} \\
B & 0 & L_{R2} & A_{2}%
\end{array}%
\right) \left(
\begin{array}{c}
\psi _{1}^{\prime } \\
\psi _{1}^{\prime \prime } \\
\psi _{2}^{\prime } \\
\psi _{2}^{\prime \prime }%
\end{array}%
\right) =\Gamma \left(
\begin{array}{c}
\psi _{1}^{\prime } \\
\psi _{1}^{\prime \prime } \\
\psi _{2}^{\prime } \\
\psi _{2}^{\prime \prime }%
\end{array}%
\right) ,  \label{4-6}
\end{equation}

\begin{equation}
\begin{split}
A_{j}& \equiv \frac{vn_{j0}n_{js,\zeta }}{2n_{js}^{2}}-\frac{vn_{j0}}{n_{js}}%
\frac{\partial }{\partial \zeta }, \\
B& \equiv -2g_{12}\sqrt{n_{1s}n_{2s}}, \\
L_{Ij}& \equiv \frac{1}{2}\frac{\partial ^{2}}{\partial \zeta ^{2}}-\frac{1}{%
2}\frac{n_{j0}v^{2}}{n_{js}^{2}}+\frac{1}{2}%
(v^{2}-p^{2})-g_{jj}n_{js}-g_{lj}n_{ls}+\mu _{j}, \\
L_{Rj}& \equiv \frac{1}{2}\frac{\partial ^{2}}{\partial \zeta ^{2}}-\frac{1}{%
2}\frac{n_{j0}v^{2}}{n_{js}^{2}}+\frac{1}{2}%
(v^{2}-p^{2})-3g_{jj}n_{js}-g_{lj}n_{ls}+\mu _{j}, \\
j& =1,2,~l=1,2,~l\neq j.
\end{split}
\label{4-7}
\end{equation}%
System (\ref{4-6}) determines the growth rates $\Gamma _{1,2}(p)$ of small
perturbations traveling along the dark-soliton's crest. The result is the
presence of two unstable branches, examples of which are shown in Fig.~\ref%
{fig5}. As one can see, both branches indeed feature regions of wave
vector $p$ with $\mathrm{Re}\,\Gamma (p)>0$. The transition to convective
instability should be considered separately for each branch.
\begin{figure}[tb]
\begin{center}
\includegraphics[width=15cm,clip]{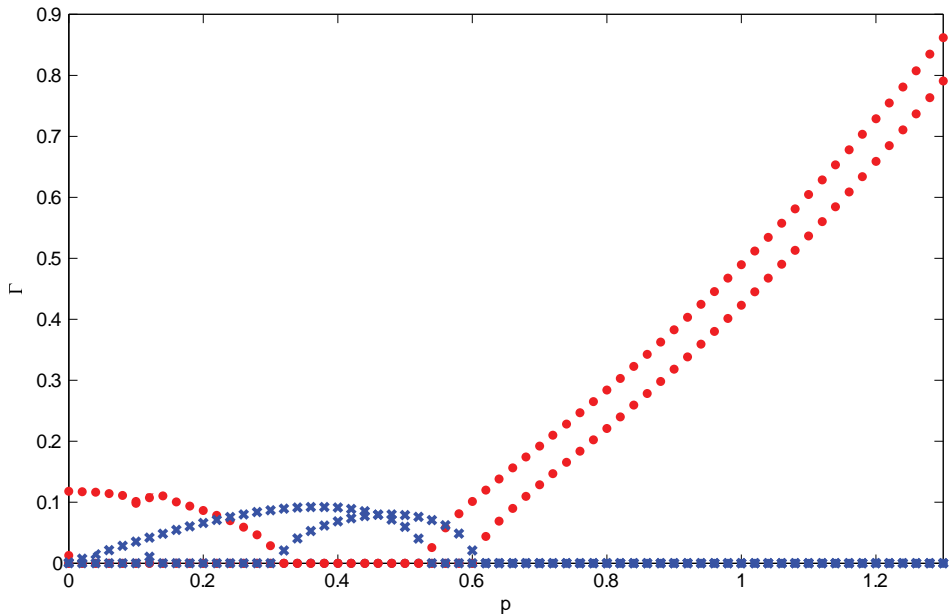}
\end{center}
\caption{(Color online.) Two branches of dispersion curves $\Gamma (p)$ for unstable
disturbances around the dark-soliton solution. Real parts of $\Gamma $ are
shown by crosses, and imaginary parts by dots. Parameters are $g_{11}=1$,$%
\,g_{22}=1.6$, $\,g_{12}=0.1$, $n_{10}=1$,$\,\ n_{20}=0.6$.}
\label{fig5}
\end{figure}

Returning to the reference system attached to the moving obstacle, we get
dispersion relations
\begin{equation}
\omega _{1,2}(p)=U_{s}p+i\Gamma (p,v),  \label{4-8}
\end{equation}%
where $U_{s}=v\sin \varphi \equiv aU/\sqrt{1+a^{2}}$ is the component of the
flow velocity along the dark soliton. The perturbations may be represented
as Fourier integrals over linear modes obeying dispersion relations (\ref%
{4-8}),
\begin{equation}
\delta n_{1,2}\propto \int_{-\infty }^{\infty }\delta \tilde{n}%
_{1,2}(p)e^{i[py-\omega _{1,2}(p)t]}dp.  \label{4-9}
\end{equation}%
This representation implies that the integral is convergent, i.e., the wave
packet is finite along coordinate $y$. Its time dependence at fixed value of
$y$ is determined by dispersion relations $\omega _{1,2}(p)$. Since
expressions (\ref{4-8}) contain imaginary parts, $\delta n_{1,2}$ can grow
exponentially with time, which implies instability of the dark soliton, as
is well known for the zero flow, $U_{s}=0$. However, it may happen that, for
$U_{s}$ large enough, wave packets are carried away so fast that they cannot
grow at fixed value of $y$, which is precisely the transition from absolute
to convective instability \cite{LL10}. Mathematically, the convective
instability means that one can transform integrals over $p$ in wave packets (%
\ref{4-9}) into integrals over $\omega $, because these wave packets have
finite duration. In other words, function $\omega =\omega (p)$ can be
inverted to define single-valued dependence $p=p(\omega )$. Therefore, the
distinction between absolute and convective instabilities depends on
analytical properties of dispersion relations (\ref{4-8}) \cite%
{LL10,sturrock}. Actually, the transition from absolute to convective
instability is determined by critical points $p_{\mathrm{cr}}$ where $%
d\omega /dp=0$, and function $p=p(\omega )$ changes its behavior: at $%
U_{s}<\left( U_{s}\right) _{\mathrm{cr}}$ it is represented in the complex $%
p $ plane by disconnected curves, whereas for $U_{s}>\left( U_{s}\right) _{%
\mathrm{cr}}$ these curves are connected with each other. In the latter
case, one can deform the contour of the integration over $p$, with regard to
the single-valuedness of $\omega =\omega (p)$, so as to transform it into an
integral over $\omega $. In other words, the spatial Fourier decomposition
of the perturbation wave packet can be transformed to a temporal form, which
means that the instability is convective.

As is known, the asymptotic behavior of integrals (\ref{4-9}) is determined
by branching points of function $p=p(\omega )$, where $d\omega /dp=0$. This
yields equation
\begin{equation}
U_{s}=-i\frac{d\Gamma }{dp},  \label{5-1}
\end{equation}%
where, as one can see in Fig.~5, $\Gamma (p,v)$ has either real or purely
imaginary values for real $p$. Therefore, critical values of $U_{s}$, at
which disconnected contours transform into connected ones, correspond to the
appearance of a double root $p_{\mathrm{br}}$ of Eq.~(\ref{5-1}) on the real
axis of $p$. This means $dp_{\mathrm{br}}/dU_{s}=\infty $ at $U_{s}=\left(
U_{s}\right) _{\mathrm{cr}}$. The differentiation of Eq.~(\ref{5-1}) with
respect to $U_{s}$ then leads to equation
\begin{equation}
\left. \frac{d^{2}\Gamma }{dp^{2}}\right\vert _{p=p_{\mathrm{cr}}}=0
\label{5-2}
\end{equation}%
for the corresponding critical value, $p_{\mathrm{cr}}(v)$. The substitution
of that value into Eq.~(\ref{5-1}) yields function $U_{s}(v)$. When this
function is known, we find, with the help of relations
\begin{equation}
v=\frac{U}{\sqrt{1+a^{2}}},\quad U_{s}=U\sin \varphi =\frac{Ua}{\sqrt{1+a^{2}}}%
=av,  \label{5-3}
\end{equation}%
the slope,
\begin{equation}
a_{\mathrm{cr}}(v)=\frac{U_{s}(v)}{v},  \label{5-4}
\end{equation}
 and velocity,
\begin{equation}
U(v)=v\sqrt{1+a_{\mathrm{cr}}^{2}(v)},  \label{5-5}
\end{equation}%
as functions of $v$ for all values in the interval $(0<v<c_{+})$. As a
result, we obtain, in a parametric form, the dependence $U(a)$ for the curve
separating regions of the absolute and convective instabilities. Two such
curves for both branches are shown in Fig.~6, where the region of the
convective instability is located above both curves.
\begin{figure}[tb]
\begin{center}
\includegraphics[width=13cm,clip]{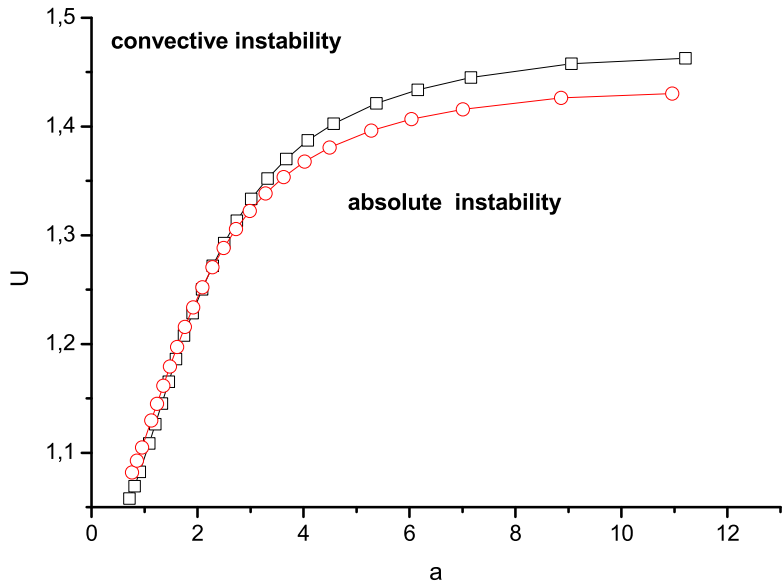}
\end{center}
\caption{(Color online.) Curves separating regions of the absolute and convective
instabilities of dark solitons in the two-component condensate with
parameters $g_{11}=1$, $g_{22}=1.6$, $g_{12}=0.1$, $n_{10}=1$,$\,\
n_{20}=0.6 $. The respective sound velocities are $c_{+}=1.03,\,c_{-}=0.95$.
Squares and circles refer to two different unstable modes.
Below each curve the corresponding mode is absolutely unstable and dark solitons
cannot be created by the flow with velocity $U$ less than $U_{cr}\cong1.5$
for most slopes $a$. Dark solitons become just convectively unstable
(effectively stable) above both curves, that is for $U>u_{cr}\cong1.5$.}
\end{figure}
It is seen that the dark solitons with large slopes $a$ undergo the
transition to the convective instability at flow velocities greater than $%
U_{cr}\approx 1.5$, as suggested also by the numerical findings reported
above.

\section{Conclusion}

In this work, we have considered the effect of dragging a supercritical
obstacle through a two-component Bose-Einstein condensate, which was
motivated by a number of recent experiments in such settings, predominantly
examining the dynamics of mixtures of hyperfine states of $^{87}$Rb. The
presence of two speeds of sound in the mixture results in the existence of
two Mach cones. The motion of the obstacle, in turn, produces two main
features, namely, the linear ``ship waves" and the oblique dark solitons.
For the former pattern, we have developed a description of their density
oscillations, which occur outside the Mach cones, in good agreement with
numerical findings. On the other hand, for the dark solitons we have
developed an analytical description of their profile, which was also found
to be in good agreement with numerical observations. In particular, it was
predicted that the dark solitons are confined to the area inside the outer
Mach cone, which was confirmed by the simulations.

This work may be a relevant starting point towards a more detailed
understanding of the interplay between the inter-species interactions and
the loss of superfluidity caused by the super-critical motion of defects.
Among natural extensions of the present setting, there may be the consideration
of the three-dimensional context, with \textit{vortex rings} being formed as
a result of the motion of the obstacle [which is closest to the experimental
settings of reported in Refs. \cite{engels07,engels08}]. Another possibility
would be to consider the supercritical motion of the obstacle in a
three-component spinor mixture \cite{dabr}, where it would be relevant to
examine the role of the spin-dependent and spin-independent parts of the
interatomic interactions in producing patterns such as those considered
herein. Such studies are currently in progress and will be reported in
future works.

\subsection*{Acknowledgments}

We appreciate a valuable discussion with L.P.Pitaevskii.
Yu.G.G. and A.M.K. thank RFFI (Russia) for financial support.
P.G.K. gratefully acknowledges support from NSF-CAREER,
NSF-DMS-0619492 and NSF-DMS-0806762,
as well as from the Alexander von Humboldt Foundation.
The work of D.J.F. was partially supported by the Special Research Account
of the University of Athens.

\end{document}